\input harvmac
\noblackbox
\def\ias{\vbox{\sl\centerline{Department of Physics, University of 
California at San Diego}
\centerline{9500 Gilman Drive, La Jolla, CA 92093 USA}}}

\lref\one{C. Callan and J. Harvey, Nucl. Phys. {\bf{B250}} (1985) 427.}
\lref\two{J. Blum and J. Harvey, Nucl. Phys. {\bf{B416}} (1994) 119.}
\lref\three{J. Blum and J. Harvey, unpublished.}
\lref\four{J. Maldacena, hep-th/9711200.}
\lref\five{T. Banks and M. Green, hep-th/9804170.}
\lref\six{J. Blum, hep-th/9712233.}
\lref\seven{M. Douglas and G. Moore, hep-th/9603167.}
\lref\eight{K. Intriligator, Nucl. Phys. {\bf{B496}} (1997) 177.}
\lref\nine{J. Blum and K. Intriligator, Nucl. Phys. {\bf{B506}} (1997) 223.}
\lref\ten{J. Blum and K. Intriligator, Nucl. Phys. {\bf{B506}} (1997) 199.}
\lref\eleven{Z. Kakushadze, hep-th/9803214; hep-th/9804184.}
\lref\thirteen{P. Aspinwall, Nucl. Phys. {\bf{B496}} (1997) 149.}
\lref\twelve{E. Witten, Nucl. Phys. {\bf{B460}} (1996) 541.}
\lref\fourteen{E. Witten, hep-th/9805112.}
\lref\fifteen{S. Kachru and E. Silverstein, hep-th/9802183.}
\lref\sixteen{O. Aharony, Y. Oz, and Z. Yin, hep-th/9803051.}
\lref\seventeen{A. Fayyazuddin and M. Spalinski, hep-th/9805096.}

\Title{\vbox{\baselineskip12pt
\hbox{UCSD/PTH 98-20}\hbox{hep-th/9806012}}}
{\vbox{\centerline{Anomaly Inflow at Singularities}}}

{\bigskip
\centerline{Julie D. Blum}
\bigskip
\ias

\bigskip
\medskip
\centerline{\bf Abstract}

Many noncompact Type I orbifolds satisfy tadpole constraints yet are
anomalous.  We present a generalization of the anomaly inflow mechanism
for some of these cases in six and four dimensions.}

\Date{5/98}

\newsec{Introduction}

The purpose of this note is to argue that the anomaly inflow mechanism \one\
in the context of gauge and gravitational defects \two\three\ can be
extended to the singular case where the scale size of these defects vanish.
We will consider noncompact type I orientifolds which are consistent 
and cancel tadpole anomalies.  In type II in six dimensions, the interchange 
of sources of curvature and sources of the field strength of the NS 
antisymmetric tensor (NS fivebranes) under T-duality makes it appropriate to
also regard gravitational defects as fivebranes.  In type IIA or M theory
the theory on the usual fivebranes is anomalous, and a current flows
onto the brane from the outside.  This current could complicate the proposed
relationship \four\ between the large $N$ theory of the M theory branes
and supergravity on $AdS_7\times S^4$.  Alternatively, one might be able
to deduce some nonperturbative correlations from the anomalous coupling 
similar to the $AdS_5$ case \five .  

In \six\ it was proposed that in some cases nonperturbative effects 
allowed one to regard singular gravitational defects of the type I $SO(32)$
theory as NS fivebranes of the heterotic $E_8\times E_8$ theory and vice
versa.  We would like to extend some of these ideas to threebranes.  In
type IIB a gravitational defect of codimension three cannot be considered
as a threebrane because there is no anomalous coupling to the Euler
characteristic.  Since there are no gauge fields, there can be no gauge
defects of the appropriate codimension.  In type I, however, we can
have gauge defects of codimension three in the presence of gravitational
defects.  The threebrane potential is generally projected out of the theory.
Nevertheless, there is the possibility of wrapping fivebranes on two-cycles.
There is, thus, the possibility of regarding this kind of singularity as
a threebrane.  We will first discuss anomaly inflow at singularities of 
codimension five.  Then we will argue, perhaps, naively that anomaly inflow
currents of gauge charge can also occur in four dimensions in the presence
of the above mentioned singularity.

\newsec{Six Dimensional Anomalies}

We will discuss here the noncompact ALE singularities in type I theory
\seven\eight\nine\ten .  It has been shown \nine\ten\ that tadpole
anomaly cancellation implies the cancellation of spacetime anomalies in
the fivebrane gauge group of these models.  We now wish to show how the
anomalies of the ninebrane gauge group are exactly those of a current 
flowing out of the ALE space and onto the singularity.  The net effect is
that the full ten-dimensional theory is not anomalous.

In the particular cases that concern us here the twelve-form for the 
ten-dimensional anomaly factorizes in the form $X_{12}=X_4 X_8$
where the six-dimensional anomaly is derived by descent from $X_8$
and the field strength of the R-R antisymmetric tensor satisfies
$dH_{RR}=X_4$ with $X_4=trR^2 -trF^2$(numerical constants are being
ignored here).  The integral of $X_4$ gives the
bulk contribution of the Euler characteristic $\chi$ minus that of
the instanton number $I$ which is integral since $H_{RR}$ is quantized.
One significant question is whether the bulk contributions
of $\chi$ and $I$ are necessarily equal if $\chi=I$.  We want to see
whether anomaly inflow can play a role in theories without physical
fivebranes.  These theories are the ${\bf Z}_{2N}$ orbifolds with
Wilson lines breaking $SO(32)$ to $SO(16)\times SO(16)$ and the 
${\bf Z}_{2N+1}$ orbifolds with Wilson lines yielding $SO(16)\times U(8)$.

The ${\bf Z}_{2N}$ orbifolds satisfy $I=\chi+F$ while the ${\bf Z}_{2N+1}$ 
orbifolds have the relation $I=\chi+F-{1\over 2N+1}$ where $F$ is 
the number of 
physical fivebranes on the Coulomb branch (which can be fractional).
By comparison with a ${\bf Z}_3$ orbifold of K3, we deduce that there is 
a ``standard embedding''  for this case which is consistent with there
being no spacetime anomaly.  This case is the only one with equal bulk
contributions of $\chi$ and $I$
that corresponds to a possible compact orbifold and the only one without
anomalies.  The bulk contribution to $\chi$ for a ${\bf Z}_N$ orbifold is
$N-1/N$ while the boundary contribution is $1/N$.  For $N>3$ the bulk
contribution to $I$ should be $F+N-8-1/N$, and the boundary contribution
should be $8+1/N$ for $N$ even and $8$ for $N$ odd to be consistent 
with anomaly cancellation.  (There are
no anomalies for $F=8$.)  It would be interesting to verify these numbers
directly.  That there is a net gravitational contribution to the $H_{RR}$
 charge
for $F=0$ is consistent with the interpretation of the singularity
as fivebranes.  When the bulk contribution of $I$ is negative ($F+N<9$),
this contribution could be interpreted as gravitational.  In this case one
expects the corrections to the gauge theory argued for in \six\ to be
important, and anomaly inflow currents to be gravitational.  Note also 
that the $H_{RR}$ charge changes sign at $F=8$.  We have not discussed the
${\bf Z}_2$ case which would require a negative boundary contribution.

Let us consider one other case.  For unbroken $SO(32)$ anomaly 
considerations
lead us to expect that the bulk contribution to $I$ will be $F+2N-24-1/N$,
and the boundary will be $24-N+1/N$ for even $N$; and the bulk will be
$F+2N-24-2/N$ and the boundary $24-N+1/N$ for 
odd $N$ which implies that one can cancel the
R-R charge in the bulk for $F=24-N$ (even $N$) and $F=24-N+1/N$ (odd $N$).
The above allows for the construction of a compact K3 for $I=24$ as expected
if the boundary is included in the compactification with 
added curvature.
For the nonabelian orbifolds without Wilson lines, anomalies also cancel
for $I=24$.
We will not try to extend this analysis to
other choices of Wilson lines but will note that determining the bulk
contributions generally will impose restrictions on the possibilities for
Wilson lines at compact K3 singularities.  In general, there will not be 
a choice of $F$ to cancel anomalies and, thus, no possible compactification.
We also note that the $H_{RR}$
charge can change integrally leaving $I$ constant, and we expect such a
change in the transition to the Higgs branch.

Now we can see how anomaly inflow works in these theories.  To cancel the
anomaly
\eqn\asix{c_6\int{d^{10} x X^1_6(\Lambda) X_4}}
we need the counterterm
\eqn\csix{c_6\int{d^{10} x H_{RR}X_7}}
where $\delta X_7=dX^1_6(\Lambda)$ with $\Lambda$ a gauge or gravitational
parameter and $c_6$ is a constant.  The integral of $H_{RR}$ over 
the boundary of the ALE space
is nonzero and quantized showing that there is a current through this
boundary with divergence compensating for the anomalous current at the 
origin given by \asix .  Our crucial assumption was that the bulk can
contribute
to the $H_{RR}$ charge beyond the effect of physical fivebranes.

\newsec{Four Dimensional Anomalies}

Let us try to extend this analysis to some noncompact orbifolds of 
codimension three.  The discussion in this section will be fairly
speculative as there do not seem to be as yet the mathematical results 
for these orbifolds similar to the ALE ones.  Because the tadpole
equations generally do not have solutions, we will confine our analysis
to ${\bf Z}_N$ orbifolds with $N$ odd.  These cases have also been discussed
by \eleven .  There are solutions of the tadpole constraints for all odd $N$,
but only the cases $N=3$ or $N=7$ which correspond to possible toroidal
compactifications give theories without nonabelian gauge anomalies in four
dimensions.  The gauge group is determined by embedding the orbifold action
into Wilson lines that break the ninebrane gauge group $SO(32)$ to a
subgroup.  There are no perturbative fivebranes.

Our speculation is that an anomaly inflow mechanism involving nonperturbative
fivebranes wrapped on two-cycles is the necessary ingredient to make
sense out of these theories.  As in six dimensions, the $H_{RR}$ charge
need not vanish since the orbifolds are noncompact.  The charge should,
however, be quantized.  Since these fivebranes are also instantons of
zero scale size \twelve\ in the ninebrane gauge group, there will necessarily
be some gauge field strengths of the ninebrane group turned on at the location
of the fivebrane.  Since the holonomy of a smooth supersymmetric $N=1$
compactification to four dimensions must be $SU(3)$, we might expect
that these instantons are embedded in $SU(3)$ subgroups with the $SU(3)$
symmetry restored when the instanton has vanishing scale size, and the cycle
also has vanishing size.  

By wrapping the fivebranes on two-cycles, we obtain a net threebrane charge
which normally does not exist in type I theory.  Assuming that the 
theories with these extra fivebranes are anomalous, there are chiral zero
modes in four dimensions that contribute to the anomaly.  Our main
assumption will be that the coupling of the four dimensional anomaly in
ten dimensions is determined by the threebrane charge induced from the wrapped
fivebranes and is equal to the bulk part of the Dirac index induced from
the fivebrane instantons.  These fivebranes are all bound to the 
gravitational defect since there are no free fivebranes in these theories.

The relevant part of the anomaly twelve-form for our consideration of the
four dimensional nonabelian gauge anomaly is 
\eqn\xtwelve{X_{12}={i\over (2\pi)^5}({1\over 720}trF^6 -{1\over 24\cdot 48}
trF^4 trR^2).}
Our next assumption is that in constructing these orbifolds there is a clean
division between the field strengths due to the induced threebranes in
the transverse dimensions and the field strengths of the four dimensional
gauge group which should vanish in the transverse dimensions.  In that 
case the twelve-form factorizes as 
\eqn\xtwelvef{X_{12}={8i\over 6(2\pi)^2} tr_G F^3\times{ 1\over (2\pi)^3 48}
(tr_{G'}F^3-{1\over 8}tr_{G'}F trR^2)}
where the first term is proportional to the anomaly polynomial for 
the four-dimensional
nonabelian gauge anomaly, and the integral of the second term denoted
$X_6$ is the bulk part of the Dirac index in six dimensions.  Here $G$
is the four-dimensional gauge group and $G'$ is the gauge group in the
transverse dimensions with expectation values induced by the wrapped 
fivebranes.

With the above assumptions, we can see how anomaly inflow works in four
dimensions.  The anomaly derived from $X_{12}$ is 
\eqn\afour{c_4\int{d^{10}x X^1_4(\Lambda) X_6}}
so we need a counterterm
\eqn\cfour{c_4\int{d^{10}x G^{G'}_5 \omega^G_5}}
where $dG^{G'}_5=-X_6$ with $G^{G'}_5=dA_4-\omega^{G'}_5$, $\delta\omega^G_5=
dX^1_4(\Lambda)$, and $c_4$ is a constant.  Here, $A_4$ is
the ten-dimensional dual of the four-form obtained by reducing the six-form
that couples to the fivebrane on a two-cycle.  We obtain a gauge invariant
five-form field strength by allowing this $A_4$ to transform under $G'$ gauge
(or gravitational) transformations and adding a $\omega_5^{G'}$ with 
$d\omega_5^{G'}=X_6$.  The variation of the counterterm under four-dimensional
gauge transformations induces a current transverse to the defect 
whose divergence cancels that of the current at the location of the 
``threebranes'' \afour .

Although we cannot analyze this issue here, it is plausible that in resolving
the singularities the threebrane charge changes integrally such that there 
are phase transitions similar to \thirteen\eight\ where the fivebrane charge 
changes.  The ``T-dual'' of this mechanism may have relevance to the 
supergravity/gauge theory correspondence \four .  In fact, fourbranes obtained
by wrapping sevenbranes on three-cycles similar to the constructions
of \fourteen\ could participate in anomaly
inflow at anomalous orbifolded orientifolds \fifteen\sixteen\eleven\seventeen .
(One could also obtain potentially anomalous axion strings by wrapping 
sevenbranes on five-cycles.)
In closing, we emphasize that we have made large assumptions in deriving 
these models of anomaly inflow that need to be studied in a more mathematical
framework.  However, these models give new life to a whole class of gauge
theories that are truly string theories.

\bigskip

This work was supported in part by a UCSD contract.

\listrefs
\end